\documentclass[12pt]{iopart}
\usepackage{iopams}
\usepackage{graphicx}

\begin{document}

\title[Multi-strange correlations]{Multi-strange baryon correlations in \emph{p+p} and \emph{d+Au} collisions at
$\sqrt{s_{NN}}$ = 200 GeV}

\author{Betty Bezverkhny for the STAR Collaboration}
%\footnote[3]{To whom correspondence should be addressed (betty.bezverkhny@yale.edu)} }
\address{  }
\address{Physics Department, Yale University, P.O.Box 208120, New Haven, CT
06520, USA}
\address{  }
\address{E-mail: betty.bezverkhny@yale.edu}

\begin{abstract}
Multi-strange baryon correlations with charged hadrons have been
measured in the \textit{p+p} and \textit{d+Au} reference systems at
$\sqrt{s_{NN}}$ = 200 GeV for future comparison with correlations in
{\it Au+Au} collisions. Weak correlation peaks are observed on both
the same side as the high $p_{T}$ ($p_{T}$ $>$ 2 GeV/c) $\Xi$ (the
trigger particle), and on the away side ($\pi$ radians away in
azimuth) in the \textit{p+p} data. A quantitative analysis requires
better statistics than is presently available. Distinct peaks are
also seen in PYTHIA simulations of \textit{p+p} collisions, but
spectra comparison between PYTHIA and data shows that PYTHIA does
not reproduce the data. A clear correlation is also present in
\textit{d+Au} data, establishing a reference foundation for a future
\textit{Au+Au} study.

\end{abstract}

%Uncomment for PACS numbers title message
%\pacs{00.00, 20.00, 42.10}

% Uncomment for Submitted to journal title message
%\submitto{Journal of Physics G}

% Comment out if separate title page not required
%\maketitle

\section{Introduction}
The STAR (Solenoidal Tracker At RHIC) Collaboration has made much
progress in understanding the medium produced at RHIC. In the past
four years STAR has shown the matter created in RHIC's most central
\textit{Au+Au} collisions to be about a hundred times denser than
normal cold nuclear matter \cite{backToBack}.  STAR has also shown
collective behaviour of this ultra-dense medium \cite{flowPaper}.
However, there remains much to understand about the properties of
the medium itself.  What are the differences in
particle-antiparticle production mechanisms in this extremely dense
environment? Given this environment, are particles that contain
strangeness produced differently than the non-strange particles?
What fraction of multi-strange baryons made in top-energy
\textit{Au+Au} collisions is produced in jets? Is that the
predominant mechanism in \textit{p+p} and \textit{d+Au} collisions?
What is the extent of the $p_{T}$ range of the particles produced
predominantly in jets? How does it compare to a \textit{Au+Au}
system, where there may be a QGP?

Multi-strange baryon correlations may help understand strange
particle production modes in central \textit{Au+Au} collisions by
measuring jet-produced strangeness. Like other high $p_{T}$ charged
particles, high $p_{T}$ multi-strange baryons are likely to be a
result of fragmentation after initial state hard scattering and thus
probe the early times of the collision volume. Hard parton
scattering will often result in back-to-back jets, which produce
particles highly correlated in the same jet cone or the others on
the opposite side. Since a complete jet reconstruction in central
\textit{Au+Au} collisions is impossible due to the high multiplicity
environment, we use statistical reconstruction of strange-particle
jets by measuring azimuthal correlations of high $p_{T}$ $\Xi$
baryons with high $p_{T}$ charged tracks on an event-by-event basis.
This method has proved effective for the back-to-back jet
suppression measurement in the \textit{Au+Au} collisions
\cite{backToBack}, and is currently being used to study
$\Lambda$+$\bar{\Lambda}$ and $K^{0}_{S}$ correlations in
\textit{Au+Au} \cite{YingHQ04}. Both \textit{p+p} and \textit{d+Au}
data are suitable as a reference for a \textit{Au+Au} study
\cite{backToBack}, as in neither collision system QGP formation is
likely. This article presents the foundation of such a reference
study.

\section{Analysis method}
For the study presented in this article 14M minimum bias
\textit{p+p} events (Year 2002 data) and
 20M minimum bias \textit{d+Au} events (Year 2003 data) are used.

The main component of the STAR detector used for this analysis is
its large Time Projection Chamber (TPC) \cite{starPRC}, which has a
2$\pi$ azimuthal coverage and pseudorapidity ($\eta$) range $<|1.5|$
. Complementing the STAR TPC, the Central Trigger Barrel (CTB)
 was also used in this study \cite{Triggers}. The tracks are
reconstructed in the TPC, while the high rate capability of the CTB
is used to ensure the correlated particles come from the same event.

One  challenge in a correlation analysis is to find enough events
where a correlation is topologically feasible. In \textit{p+p}
collisions, the physics of the reaction and the efficiency of the
detector allow for topological reconstruction
\cite{multiStrangePaper} of only approximately three $\Xi^{-}$ or
$\overline{\Xi}^{+}$ per $10^{3}$ events . Only a fraction of these
trigger $\Xi$ particles have $p_{T}$ greater than the minimum
cut-off used in this study: 2 GeV/c per particle.  The low mean
multiplicity of a \textit{p+p} event (5.5 tracks) makes it difficult
to find suitable correlation partners, therefore the associated
particle cut-off is set low at 1.5 GeV/c.

To make a correlation between a $\Xi$ baryon and a high $p_{T}$
track in a given event one first applies loose selections to find
potential $\Xi^{-}$ and $\overline{\Xi}^{+}$ particles in each
event. Secondly, one looks for single tracks that would pass a given
$p_{T}$ cut-off. The background decreases as the $p_{T}$ cut-off is
raised.  After a list of eligible tracks and $\Xi$ candidates is
made, each $\Xi$ is used as a trigger particle for correlations.
Each potentially correlated high $p_{T}$ track is checked to ensure
it is not a decay product of the $\Xi$. The azimuthal angle
($\Delta\phi$) between the $\Xi$ and a high $p_{T}$ track is
calculated and plotted. After analyzing the entire event sample, the
correlation function is normalized by the number of trigger
particles and fit with two Gaussians constrained to wrap around
$2\pi$.  The means are fixed at $\Delta\phi$ = 0 and $\Delta\phi$ =
$\pi$, and the signal is assumed to sit on a flat background. A more
sophisticated background subtraction is planned for the future.

\begin{figure}[t!]
\begin{minipage}[t]{8cm}
\centering
  % Requires \usepackage{graphicx}
  \includegraphics[width=0.9\textwidth]{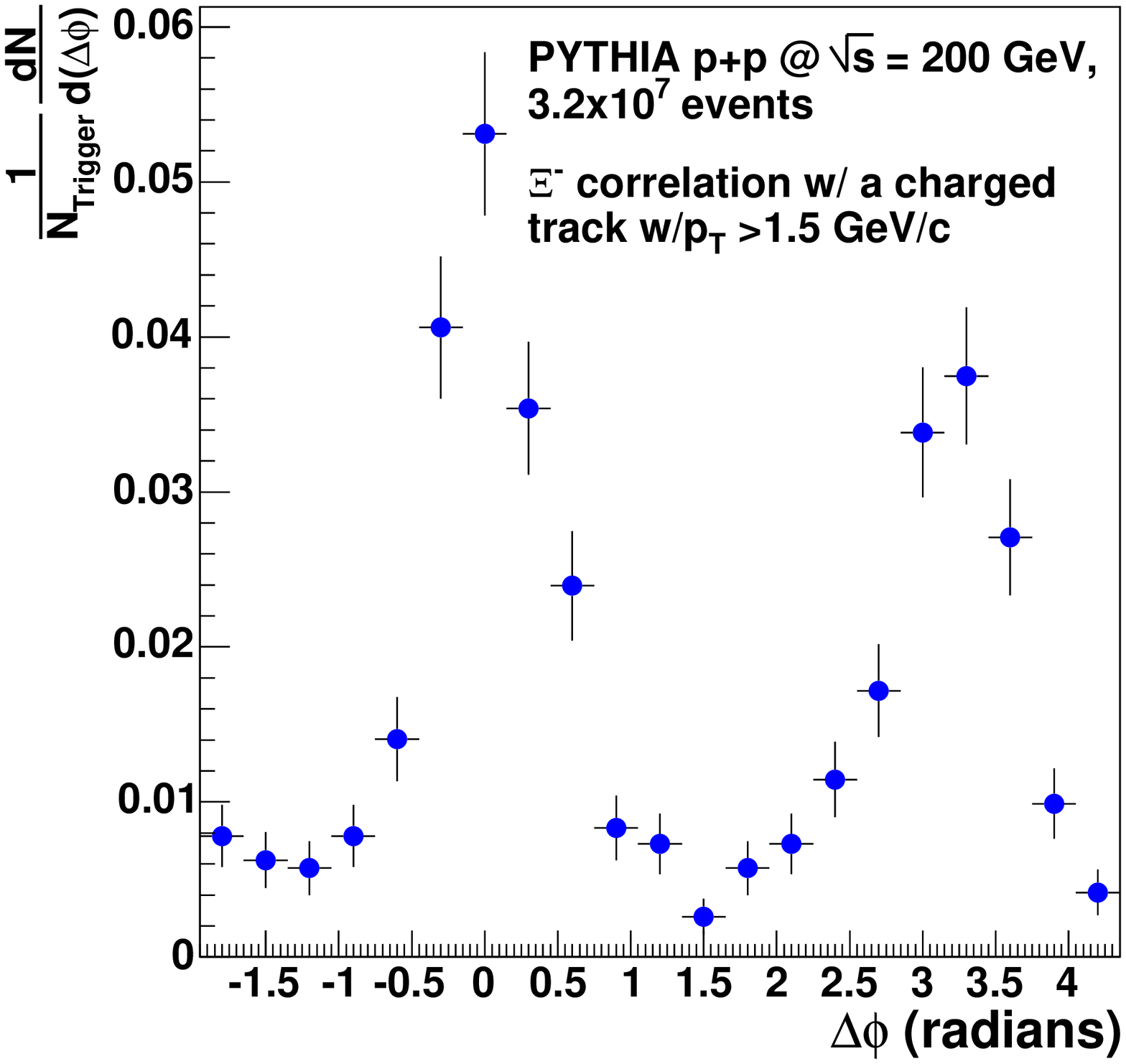}\\
  \caption{PYTHIA-simulated $\Xi^{-}$-charged primary track correlation.}\label{PythiaCorr}
%\end{figure}
\end{minipage}
\hfill
%\begin{figure}
\begin{minipage}[t]{8cm}
\centering
  % Requires \usepackage{graphicx}
  \includegraphics[width=0.9\textwidth]{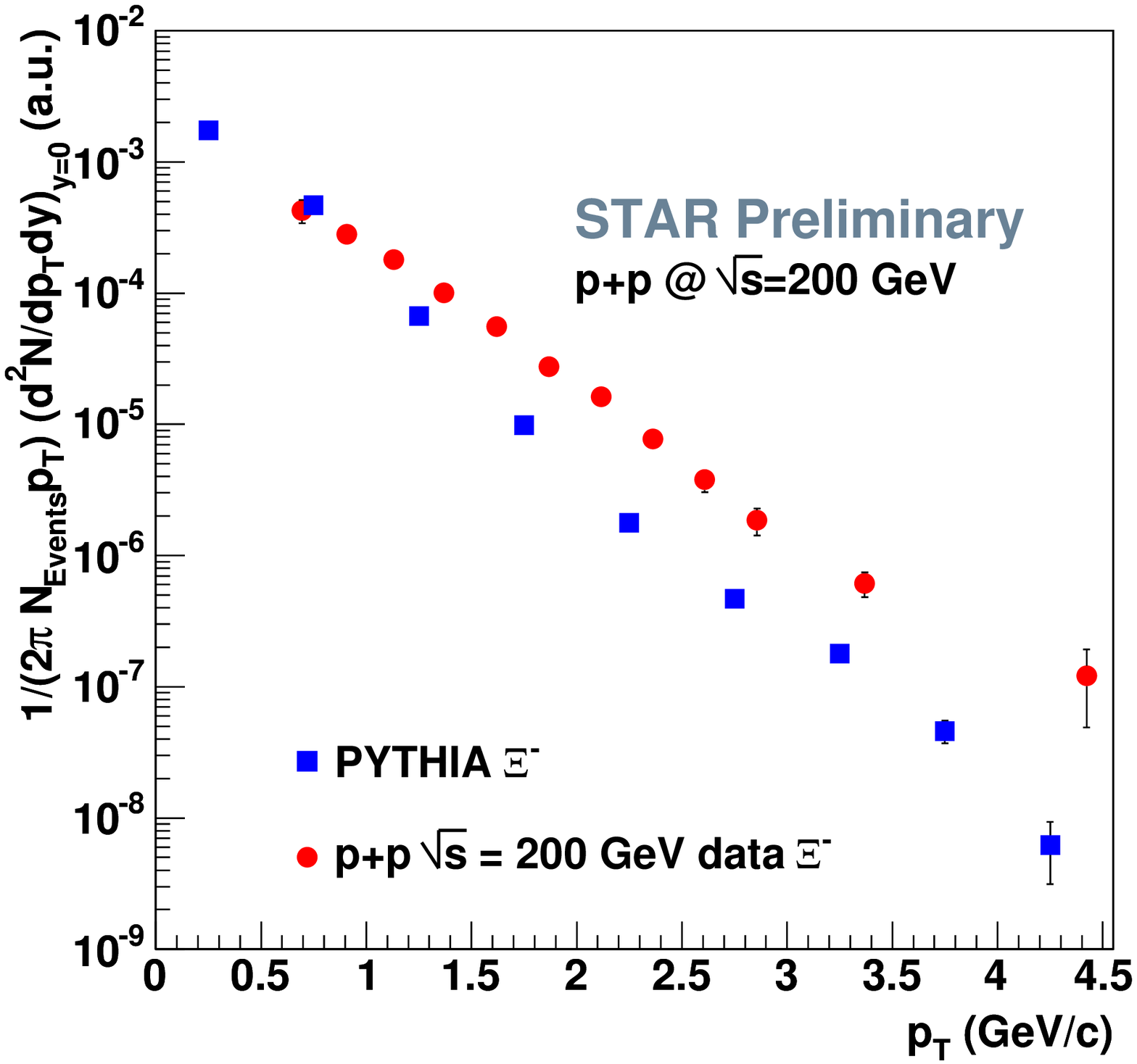}\\
  \caption{$\Xi^{-}$ spectra from PYTHIA and $\sqrt{s}$ = 200 GeV \textit{p+p} data.}\label{ppSpectra}
\end{minipage}
\end{figure}

\section{Current results}

\subsection{p+p Year 2002 data set and PYTHIA simulation}
The 2002 data set is the only sufficiently large \textit{p+p}
minimum bias data set taken so far by STAR.  The relatively clean
environment of a \textit{p+p} collision allows us to examine these
data and determine whether there are sufficient statistics to
establish a reference with which to compare central \textit{Au+Au}
collisions.

One of the ways to understand the data at hand is to perform a
simulation. This was done using the PYTHIA 6.22 event generator
\cite{PythiaMan}. $3.2\times10^{7}$ PYTHIA events were produced of
which $4.2\times10^{5}$ events had at least one $\Xi^{-}$ and
$1.9\times10^{3}$ had at least one $\Xi^{-}$ within $y < |0.75|$ and
with $p_{T}$ $>$ 2 GeV/c. Performing the identical analysis
described above for the data, one finds that with 1921 $\Xi$ trigger
particles there are 705 $\Xi$-charged hadron correlations, with 1.4
correlations per correlated particle (Table~\ref{myTable}). In other
words, if a $\Xi^{-}$ was correlated, more than a third of the time
it correlated with more than one track in the same event. The
resultant correlation is shown in Fig.~\ref{PythiaCorr}, where a
clear same-side and a distinct away-side peak are seen.  The
same-side peak obtained in this simulation contains 12\% more
correlations than the away-side peak, and is 33\% higher.

Although correlations are observed in both the \textit{p+p} data and
that of simulated particles from PYTHIA, a comparison of the two
spectra shows that the simulation does not reproduce the data. While
the PYTHIA integrated yield is higher (dN/dy = 0.00318 for PYTHIA,
dN/dy = $0.00181\pm0.00008$ the \textit{p+p} data
\cite{richardSQM}), it grossly underpredicts the yields in the
region of interest ($p_{T}>2$ GeV/c), at the same time overstating
the yields below $p_{T}$ = 0.8 GeV/c (Fig.~\ref{ppSpectra}). This
might be altered by adjusting various PYTHIA parameters, such as
tuning the hard processes parameters and allowing for parton
rescattering, or making NLO pQCD calculations.

\begin{figure}[t!]
\begin{minipage}[t]{8cm}
\centering
  % Requires \usepackage{graphicx}
  \includegraphics[width=0.9\textwidth]{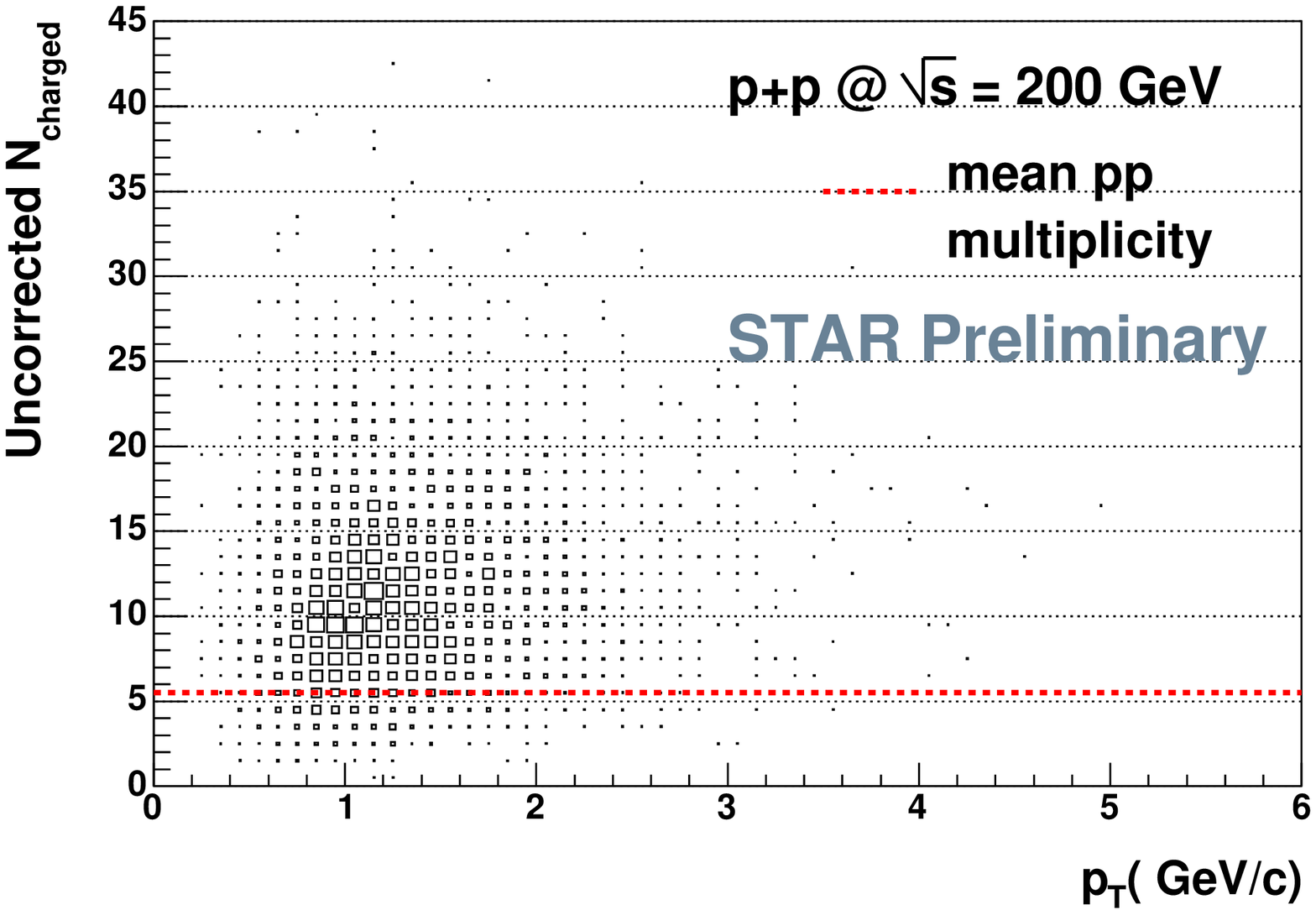}\\
  \caption{Reference event charged particle multiplicity as a function of $p_{T}$ of the $\Xi$
  in $\sqrt{s}$ = 200 GeV \textit{p+p} collisions.}\label{RefMultVsPt}
%\end{figure}
\end{minipage}
\hfill
%\begin{figure}
\begin{minipage}[t]{8cm}
\centering
 \includegraphics[width=0.9\textwidth]{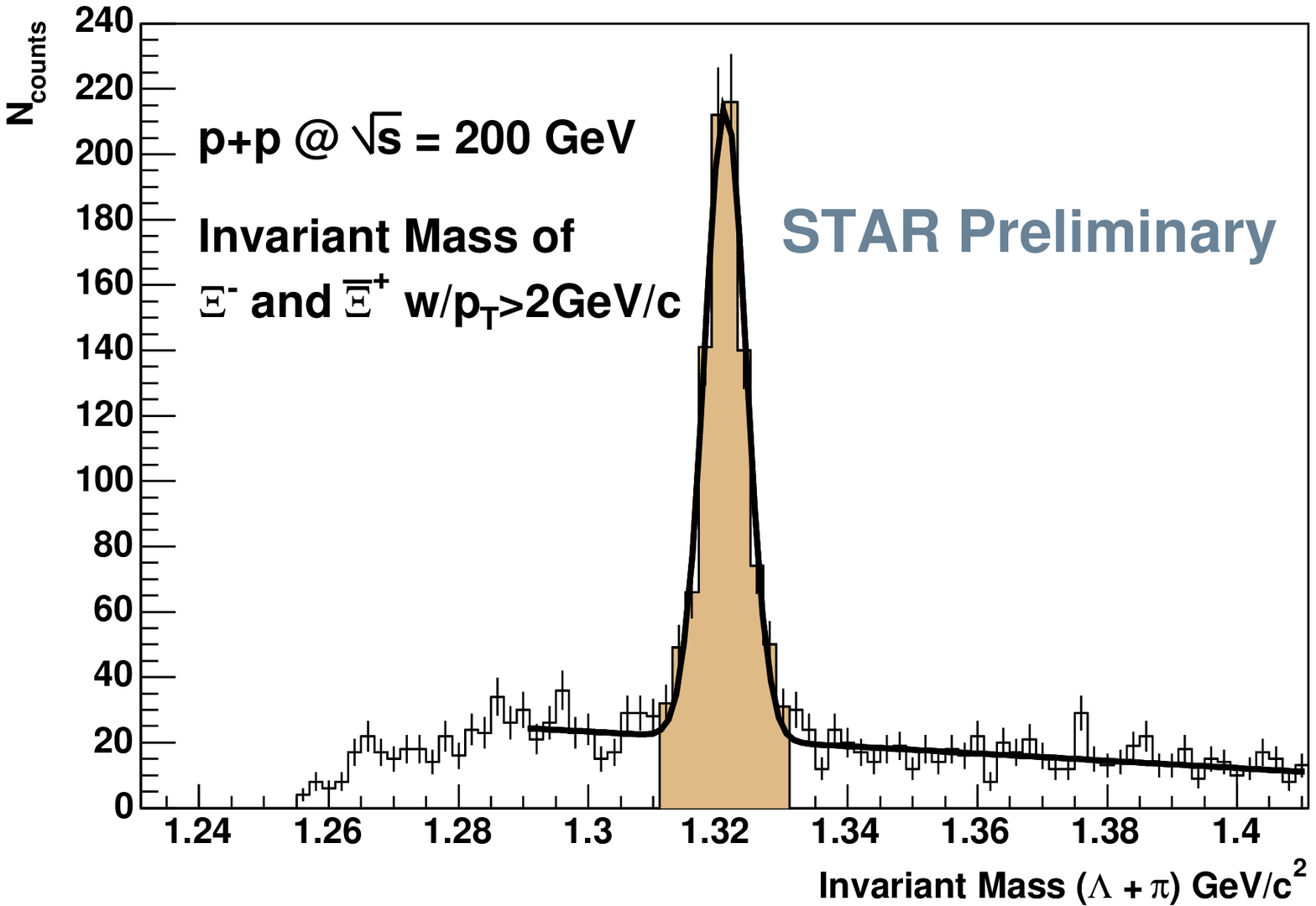}\\
  \caption{$\Xi^{-}$ and $\overline{\Xi}^{+}$ mass peak in $\sqrt{s}$ = 200 GeV \textit{p+p} data set for
  particles with $p_{T}> 2$ GeV/c.}\label{ppMassPeak}

\end{minipage}
\end{figure}

After performing a simulation, we proceed to examine characteristics
of events containing a $\Xi$ to learn about  the environment for
production of high $p_{T}$ $\Xi$ baryons in elementary processes.
One characteristic is the number of valid primary tracks found in an
event (event multiplicity).  A valid primary track in this study is
a track reconstructed with more than 15 (out of possible 46) fit
points, and which comes within 3 cm of the primary collision vertex.
Preliminary studies indicate that indeed, the multiplicity of an
event with a reconstructed $\Xi$ particle differs significantly from
that of an average minimum bias sample event. As can be seen in
Fig.~\ref{RefMultVsPt}, the mean multiplicity of an event with a
$\Xi$ produced and reconstructed is almost twice as high as the mean
multiplicity of 5.5 primary tracks per event. This enhanced
multiplicity suggests a higher mean $p_{T}$ for these events, and
thus the multi-strange particles we see in \textit{p+p} collisions
are likely to be created in more violent collisions and thus in jet
events \cite{jetSummary}. Further investigation is under way.

\begin{figure}[t!]
\begin{minipage}[t]{8cm}
\centering
  % Requires \usepackage{graphicx
   \includegraphics[width=0.9\textwidth]{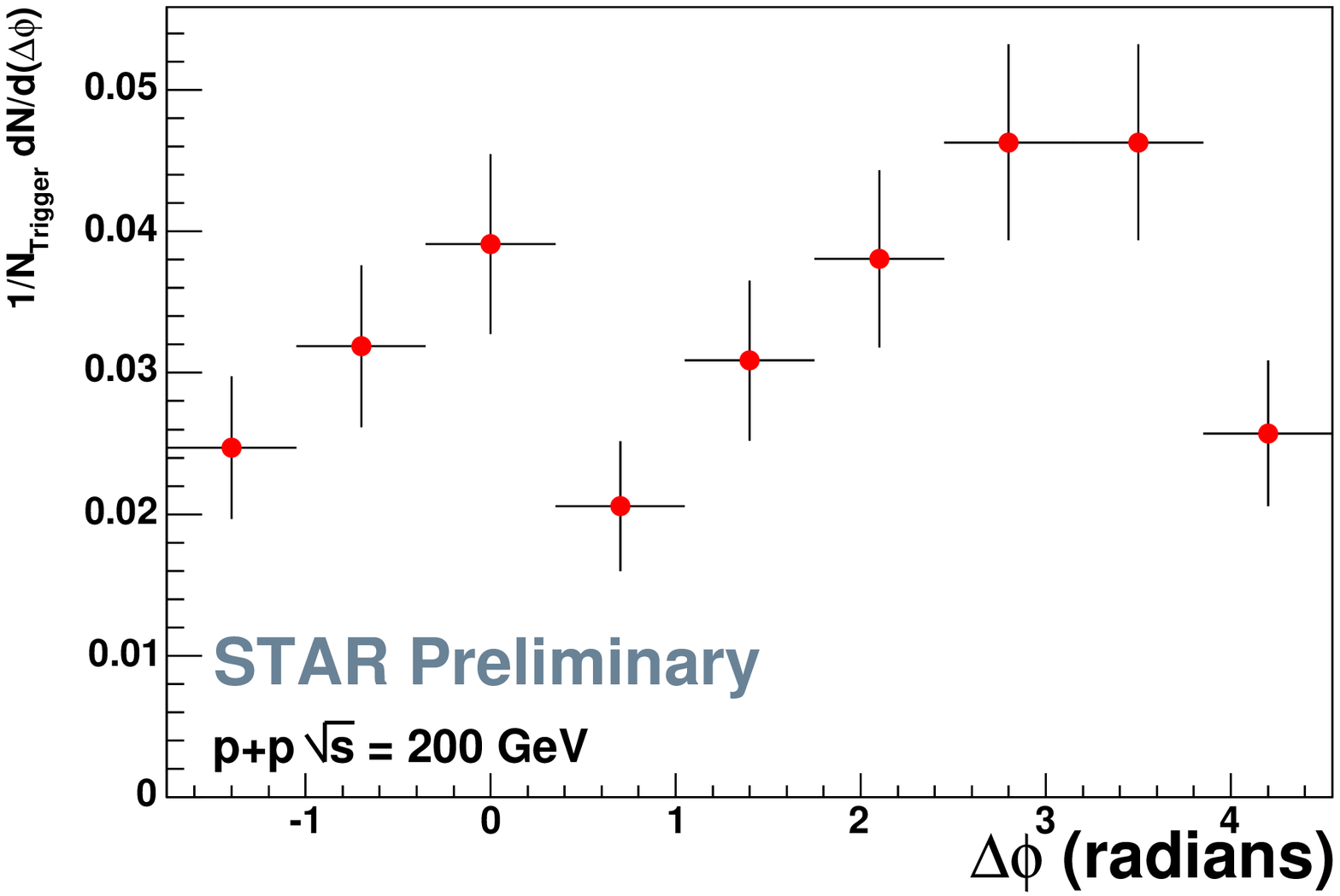}\\
  \caption{$\sqrt{s}$ = 200 GeV \textit{p+p} data uncorrected $\Xi$ - charged primary track azimuthal correlation.}\label{ppCorrelation}
%\end{figure}
\end{minipage}
\hfill
%\begin{figure}
\begin{minipage}[t]{8cm}
\centering
  % Requires \usepackage{graphicx}
  \includegraphics[width=0.9\textwidth]{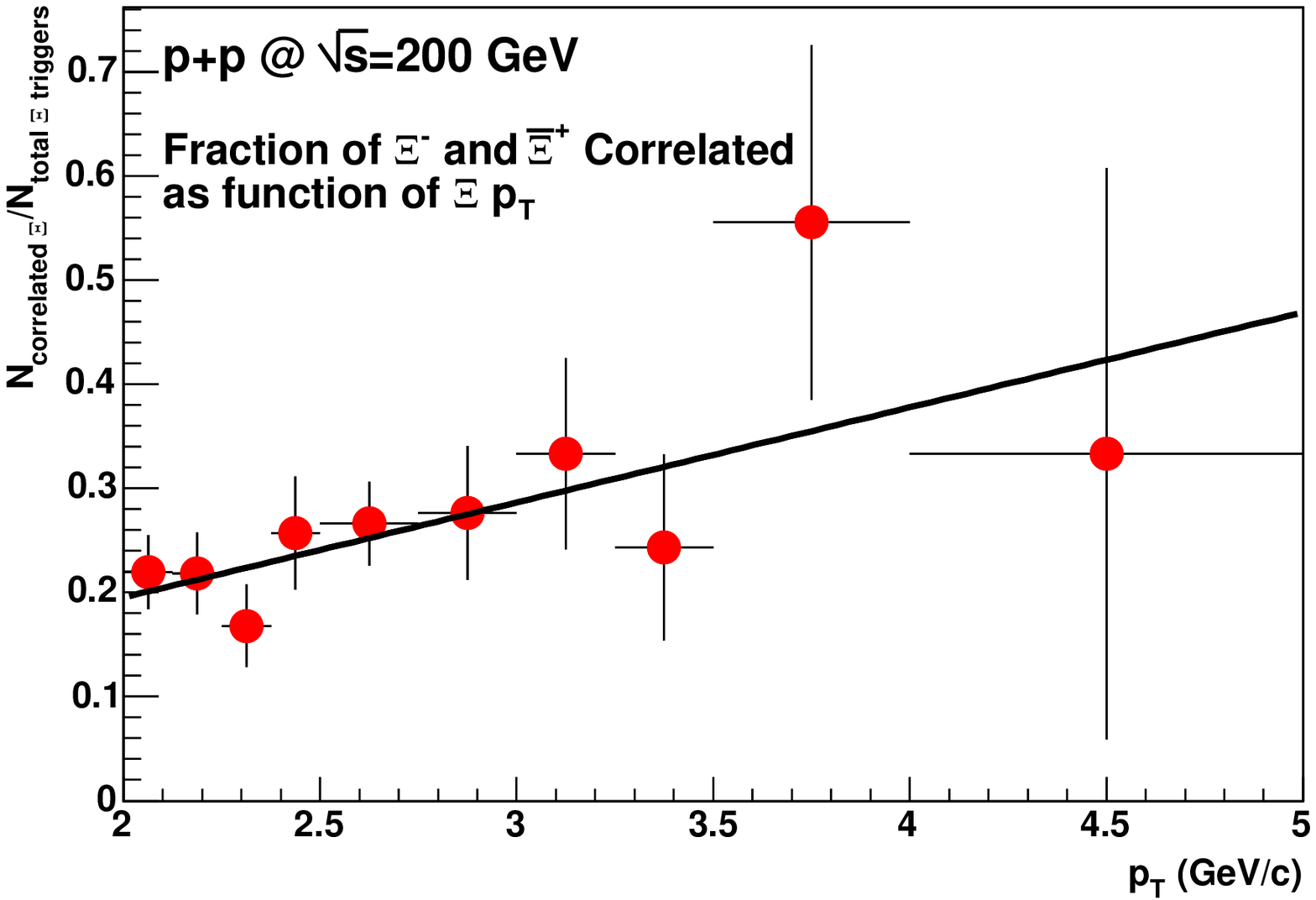}\\
  \caption{Fraction of trigger $\Xi$ particles correlated as a
  function of trigger $p_{T}$ in
  $\sqrt{s}$ = 200 GeV \textit{p+p} data set.}\label{ppPtDist}
\end{minipage}
\end{figure}

\subsection{p+p analysis results}
Loose geometrical and $\Lambda$ mass cuts were applied to find both
the $\Xi^{-}$ and its antiparticle in the minimum bias \textit{p+p}
data set. The looseness of the cuts for the selected $p_{T}$ range
allows for a 10\% increase in reconstruction efficiency compared to
cuts applied to the entire $\Xi$ $p_{T}$ range. The drawback of
loosening the selections is the slight increase in background (B)
under the signal (S) peak, as seen in Fig.~\ref{ppMassPeak}. The S/B
for the resultant peak is found to be 4.6.

A tight cut around the $\Xi$ mass peak between 1.312 GeV/c$^{2}$ and
1.330 GeV/c$^{2}$ selects the $\Xi$ candidates for correlation.
Fitting the signal with a Gaussian and a constant background yields
S = 772$\pm$31 and B = 168. Since the number of counts in the
selected mass region varies slightly from fit values, the actual
number of trigger particles was 972. Only 232 of these were
correlated resulting in 295 correlations, or 1.3 correlations per
$\Xi$ candidate, i.e., as in PYTHIA, in about a third of the events
where a correlation was possible, more than one track with $p_{T}$
$>$ 1.5 GeV/c was found (Table~\ref{myTable}).

The correlation function may be seen in Fig.~\ref{ppCorrelation}.
Although both the same- and the away-side peaks are visible, the
statistics to extract any quantitative information are lacking.
There are correlations in \textit{p+p}, and the higher the trigger
momentum, the larger fraction of the available $\Xi$ baryons is
correlated, as may be inferred from Fig.~\ref{ppPtDist}.

\begin{table}[t!]
\caption{\label{myTable}PYTHIA, \textit{p+p}, and \textit{d+Au}
values extracted: signal (a), background (b), number of trigger
particles (c), correlations obtained (d), number $\Xi$ correlated
(e), correlations per correlated $\Xi$ (f), same-side peak yield
(g), and the away-side yield (h).}
\begin{indented}
\item[]\begin{tabular}{@{}lllllllll}
  \hline
  % after \\: \hline or \cline{col1-col2} \cline{col3-col4} ...
  ~& a & b  & c & d & e & f & g & h\\
  \hline
  \textbf{PYTHIA }     & - & - & 1921 & ~705 & ~516 & 1.4 & ~394 & ~311\\
  \textbf{\textit{p+p} data}  & $~772\pm31$ & 168 & ~972 & ~295 & ~232 & 1.3 & ~- & ~- \\
  \textbf{\textit{d+Au} data} &$3576\pm66$ & 711 &4395& 4309 & 2521 & 1.7 & 2032 & 2277 \\
  \hline
\end{tabular}
\end{indented}
\end{table}

\subsection{d+Au analysis results}
\begin{figure}
\begin{minipage}[t]{8cm}
\centering
  % Requires \usepackage{graphicx}
  \includegraphics[width=0.9\textwidth]{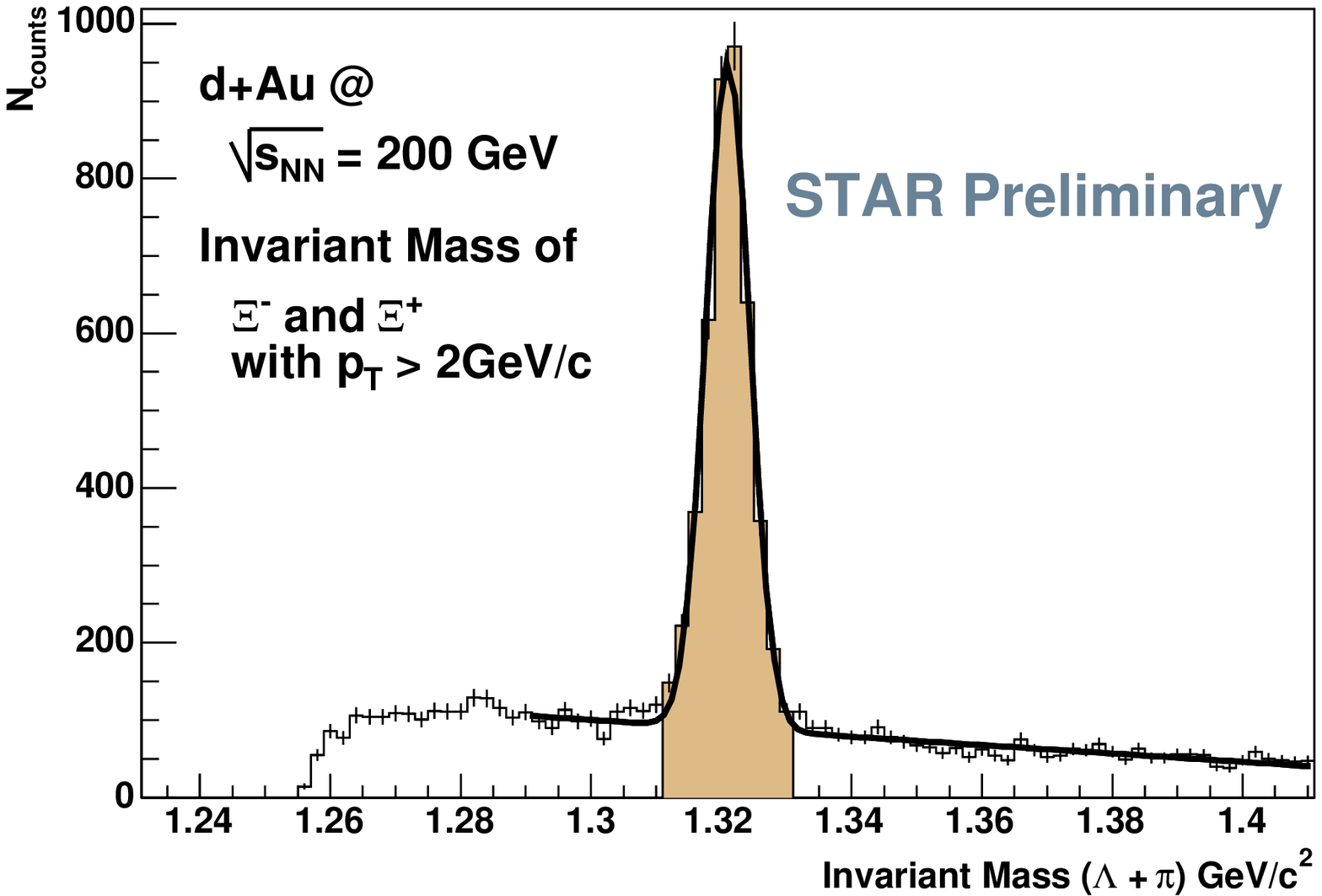}\\
  \caption{$\Xi^{-}$ and $\overline{\Xi}^{+}$ mass peak in $\sqrt{s_{NN}}$ = 200 GeV \textit{d+Au} minimum bias data set for
  particles with $p_{T}> 2$ GeV/c.}\label{dAuMassPeak}
\end{minipage}
%\end{figure}
%\begin{figure}
\hfill
\begin{minipage}[t]{8cm}
\centering
  % Requires \usepackage{graphicx}
  \includegraphics[width=0.9\textwidth]{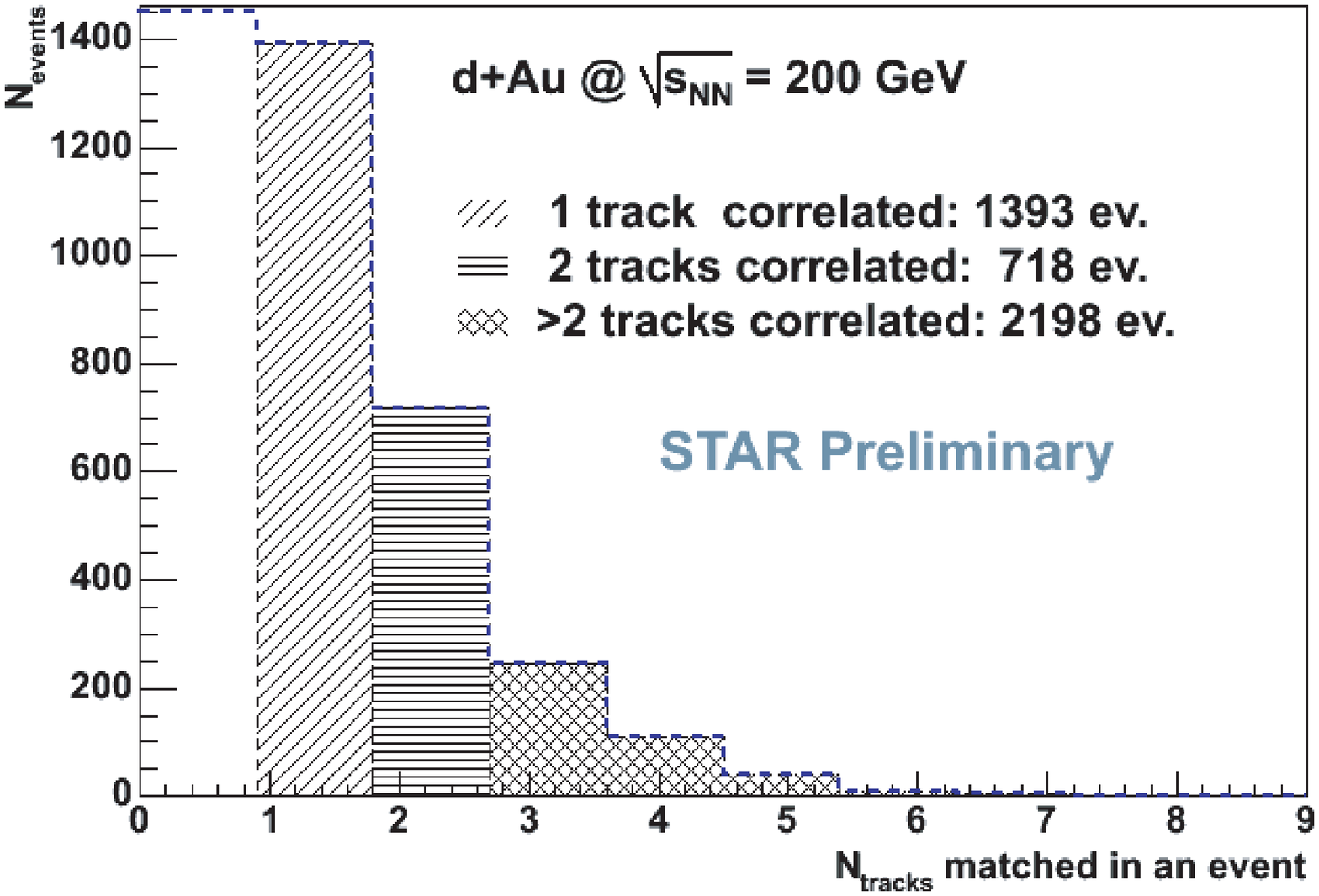}\\
  \caption{Number of uncorrected charged tracks correlated per $\Xi$ in an eligible
  $\sqrt{s_{NN}}$~=~200~GeV \textit{d+Au} minimum bias event.}\label{dAuTracksCorrel}
\end{minipage}
\end{figure}

Another \textit{Au+Au} reference to consider is the \textit{d+Au}
data set taken by STAR in 2003.  Since the collisions are no longer
nucleon-on-nucleon, but rather nucleus-on-nucleus, nuclear effects
such as the Cronin effect \cite{GVWZ}, initial state shadowing
\cite{GVWZ}, and re-scattering are present.  The \textit{d+Au}
collision environment is not as clean as in \textit{p+p} collisions;
however, the statistics are much more abundant.

\begin{figure}
\centering
  % Requires \usepackage{graphicx}
  \includegraphics[width=0.5\textwidth]{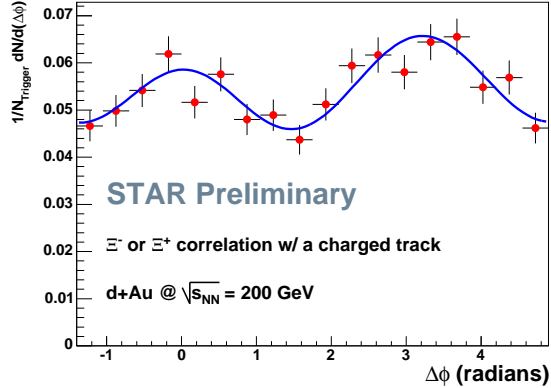}\\
  \caption{$\Xi$ - uncorrected charged primary track azimuthal correlation in \textit{d+Au} $\sqrt{s_{NN}}$~=~200~GeV minimum bias data.}\label{dAuCorr}
\end{figure}

Utilizing these statistics, and applying a tighter set of cuts than
the one used in the \textit{p+p} data set, a mass peak with over
$4\times10^{3}$ correlation candidates is obtained, as shown
Fig.~\ref{dAuMassPeak}.  As before, there is a 2 GeV/c transverse
momentum cut applied and a Gaussian plus a constant are fit to yield
the values in Table~\ref{myTable}. As demonstrated in
Fig.~\ref{dAuTracksCorrel}, only 32\% of correlated events had one
primary track with sufficiently high $p_{T}$, the others had two or
more tracks available for correlation.  This is not surprising,
since the mean multiplicity of a \textit{d+Au} event is several
times higher than that of a \textit{p+p} collision. Further study of
the multiplicity dependence of selected $\Xi$ events is planned.

Finally, we come to the correlation function in \textit{d+Au}.  As
seen in Fig.~\ref{dAuCorr}, the $d$+$Au$ data set has sufficient
statistics to fit two emerging peaks.  Contrary to what we saw in
\textit{p+p} PYTHIA simulations, and in line with the \textit{p+p}
data, the same-side peak is equal or smaller than the away-side
peak, which could be explained by depletion of the available high
$p_{T}$ tracks on the same side by the $\Xi$ decay itself, which
uses up at least 3.3 GeV of the available jet-cone energy. The peaks
in the \textit{d+Au} correlation function have widths:
$\sigma_{same}$ = 1.29 $\pm$ 0.26 radians and $\sigma_{away}$ = 1.02
$\pm$ 0.23 radians, comparable to those in \textit{p+p} data.

\section{Conclusions}
\subsection{What has been learned so far}

Despite the lack of available statistics in the current \textit{p+p}
data set, there is still information to be extracted.  We have
demonstrated a correlation between the multiplicity of a collision
and production of a $\Xi$ (Fig.~\ref{ppCorrelation}).  We also know
that the higher the trigger $\Xi$ $p_{T}$, the more likely one is to
find correlation partners for the trigger particle
(Fig.~\ref{ppPtDist}), and once an associated particle is found, it
is probable that there is more than one such particle available. All
of this combined indicates that in \textit{p+p} collisions high
$p_{T}$ $\Xi$ baryons are likely to be produced in jets.

The \textit{d+Au} data set looks very promising as a reference.  A
careful analysis of the background and setting the $p_{T}$ cut-off
higher and at multiple levels for both the trigger and the
associated particles is necessary for establishing a reliable base
line.  So far the height of the same-side peak (Fig.~\ref{dAuCorr})
is consistent with that of $\Lambda$ particles in a \textit{Au+Au}
analysis \cite{YingHQ04}, however, its size and width in relation to
the away side peak differs both from the PYTHIA predictions
(Fig.~\ref{PythiaCorr}) and the peak measured in charged-hadron
analysis \cite{backToBack}.  As is the case with high-$p_{T}$ $\Xi$
and $\overline{\Xi}^{+}$ particles in the \textit{p+p} data set,
high-$p_{T}$ $\Xi$ and $\overline{\Xi}^{+}$ particles found in
\textit{d+Au} are likely to be produced in jets.

\subsection{Outlook}
To obtain a tool for further understanding of the multi-strange
production mechanism, PYTHIA simulations need to be tuned.
Furthermore, in the \textit{d+Au} data set there needs to be a
soft-physics subtraction to understand better the background.  To
gain in statistics for the continued study, the symmetrical Gaussian
peaks can be folded \cite{Kirill}. There are also high
$p_{T}$-triggered \textit{d+Au} and \textit{p+p} data sets, yet to
be analyzed. Comparing triggered results to those in minimum bias
collisions should lead to a better understanding of the relationship
between jets and multi-strange particles.

Along with the statistics-rich $\sqrt{s_{NN}}$ = 200 GeV
\textit{Au+Au} data set taken by STAR in 2004 and soon to be
processed, there is also a new $\sqrt{s_{NN}}$ = 62 GeV
\textit{Au+Au} data set obtained by STAR during the same run year.
Looking for multi-strange correlations there will help establish the
framework for analysis in $\sqrt{s_{NN}}$ = 200 GeV \textit{Au+Au}
data.

One hopes this study, in conjunction with other identified particle
studies, builds a foundation for a better understanding of
production mechanisms for all strange particles.

%\clearpage
\end{document}